\documentclass[12pt,draftcls,peerreview,onecolumn]{IEEEtran}

\usepackage{amsmath,amssymb,graphicx,epstopdf,cite,subfigure,epsf}
\usepackage{hhline}
\usepackage{pdfpages}    
\usepackage{psfrag}
\usepackage{amssymb}
\usepackage{amsmath}      
\usepackage{bbm}
\usepackage{dsfont}   
\usepackage{pifont} 
\usepackage{cite}
\usepackage{graphics} 
\usepackage{epsfig}
\usepackage{amscd}
\usepackage{accents}
\usepackage{multirow}       
\usepackage{epstopdf}
\usepackage{algorithm,algpseudocode}
\usepackage{color}
\usepackage{varwidth}    
        
\newlength{\dhatheight}

\title{Deep Learning-based Beam Tracking for Millimeter-wave Communications under Mobility
}

\author{Sun Hong Lim,
        Sunwoo Kim, 
        Byonghyo Shim,
        and~Jun Won Choi% <-this % stops a space
\thanks{Sun Hong Lim, Sunwoo Kim, and Jun Won Choi are with Department
of Electrical Engineering, Hanyang University, Seoul, Korea,
 e-mail: shlim@spa.hanyang.ac.kr, remero,junwchoi@hanyang.ac.kr.}% <-this % stops a space
\thanks{Byonghyo Shim is with Department
of Electrical and Computer Engineering, Seoul National University, Seoul, Korea, e-mail: bshim@snu.ac.kr.}% <-this % stops a space
}

\begin{document}

% make the title area
\maketitle
       
\begin{abstract}
In this paper, we propose a deep learning-based beam tracking method for millimeter-wave (mmWave) communications. Beam tracking is employed for transmitting the known symbols using the {\it sounding beams}  and tracking time-varying channels to maintain a reliable communication link. When the pose of a user equipment (UE) device varies rapidly, the mmWave channels also tend to  vary fast, which hinders seamless communication. Thus, models that can capture temporal behavior of mmWave channels caused by the motion of the device are required, to cope with this problem. Accordingly, we employ a deep neural network to analyze the temporal structure and patterns underlying in the  time-varying channels and the signals acquired by inertial sensors. We propose a model based on long short term memory (LSTM) that predicts the distribution of the future channel behavior based on a sequence of input signals available at the UE. This channel distribution is used to 1) control the sounding beams adaptively for the future channel state and 2) update the channel estimate through the {\it measurement update step} under a sequential Bayesian estimation framework. Our experimental results demonstrate that the proposed method  achieves a significant performance gain over the conventional beam tracking methods under various mobility scenarios.      

\end{abstract}


\IEEEpeerreviewmaketitle   

%================================      
\section{Introduction}  
Millimeter wave (mmWave) communication has attracted significant attention for achieving the continuously increasing data throughput requirement of advanced wireless systems \cite{mmwave, mmwave2, mmwave_chan}. However, several challenges should be addressed to enable seamless communication over mmWave-band channels. In particular, the received signal power of mmWave communication systems experiences significant attenuation.  
A potential solution  is to employ directional transmit (Tx) and receive (Rx) beamforming antennas, which direct highly directional beams  in the desirable directions to enhance the signal power. Such beams are formed by appropriately adjusting the phase and amplitude of the signal for each antenna element \cite{beamforming, beamforming2}. 

Consider a base-station (BS) equipped with $N_b$ antennas and a user equipment (UE) with $N_m$ antennas. In down-link scenarios, the BS uses a beamforming vector to transmit the  data symbols to the UE and the UE applies a combining vector to receive the transmitted data  symbols. These beamforming and combining vectors determine the directions of the beams, which should be chosen to maximize the data throughput. The channel state information should be known to both the BS and UE to determine the directions of the beams. The procedure for acquiring the channel information using pilot symbols is called {\it beam training} \cite{overview,chan_est}.
In beam training, the pilot symbols are transmitted using specifically designed  Tx and Rx beams. These beams are often called {\it sounding beams} \cite{sounding}. The BS and the UE use the combinations of $M_b$ Tx sounding beams and $M_m$ Rx sounding beams to obtain the channel information. Consequently, the $M_b \cdot M_m$ pilot symbols are transmitted. Given the absence of prior knowledge about the channel, both $M_b$ and $M_m$ should be sufficiently large to cover a wide range of directions.

\begin{figure}[t]
    \centering
%\includegrachics{figures/beamtrackingprotocolfinal.eps}
    \includegraphics[width=140mm]{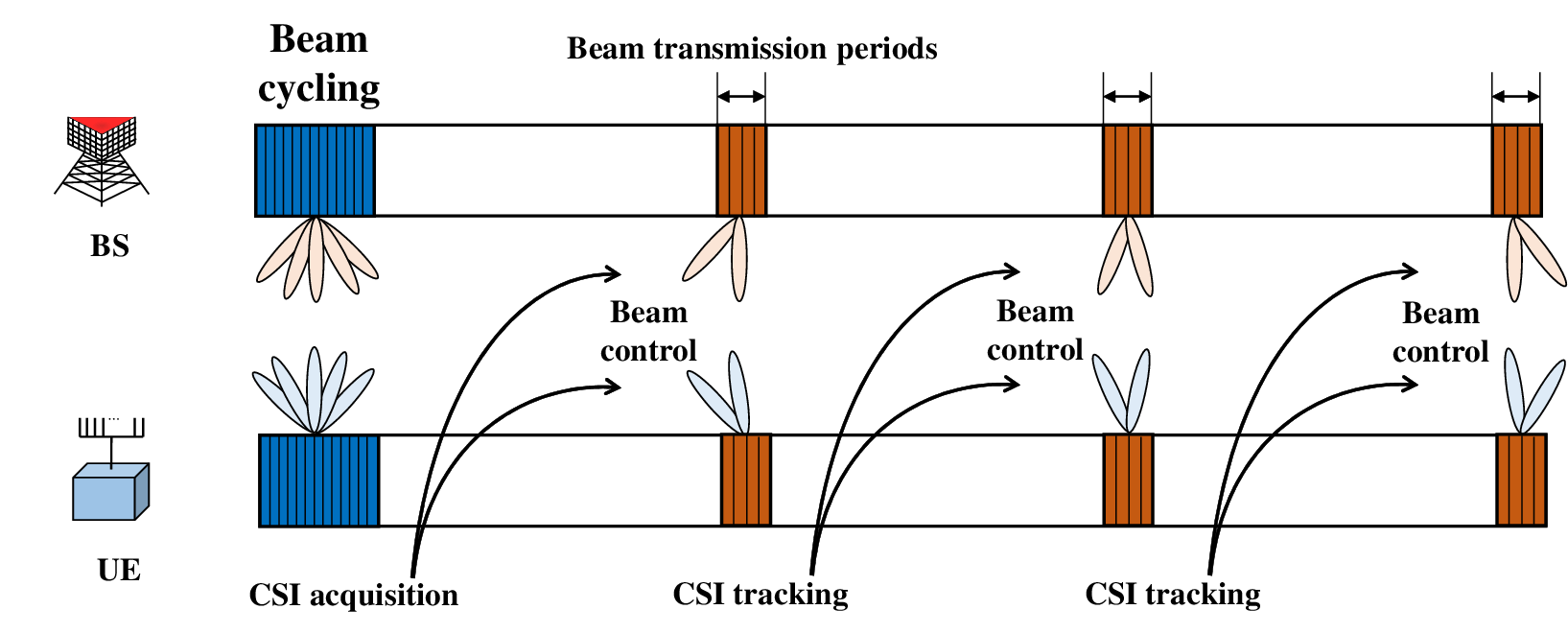}
    \caption{Illustration of beam tracking protocol}
    \label{fig:beamtracking_protocol}
\end{figure}

{\it Beam tracking techniques} have been proposed to reduce the amount of radio resources required for beam training. When the beam tracking is enabled, the BS transmits the pilot symbols using  fewer sounding beams after the initial acquisition step. (see Fig. \ref{fig:beamtracking_protocol}.) The number of sounding beams can be reduced without significantly by exploiting the temporal channel correlation. The BS and UE can use the information on the angles of arrival (AoAs) and  angles of departure (AoDs) obtained in the previous beam transmissions to direct only a few beams toward the directions that ensure good channel estimation performance.  

Two key design issues exist in implementing beam tracking systems. First, both Tx  and Rx sounding beams should be determined  in response to the time-varying AoAs and AoDs. Note that such beam control should be predictive to steer the sounding beams toward the future channel state in advance. Second,  the UE needs to update the channel estimate by using the received pilot symbols and exploiting the temporal channel correlation.

Various channel tracking methods have been proposed thus far.  
In \cite{sounding_beam}, the authors proposed a beam tracking algorithm that exploits  the temporal correlation between AoDs and AoAs. However, adapting to rapid channel variations was challenging, as omni-directional training beams were used.  
In \cite{ekf,ekf_heath,ukf,tracking_ukf,skim},  various types of Kalman filters were employed to track time-varying channels. 
In \cite{pomdp}, an optimal beam training protocol design scheme was derived based on the partially observable Markov decision process framework. 
Compressed sensing (CS) recovery algorithms \cite{cs_shim} were also used to estimate the AoDs and AoAs of multi-path channels in \cite{mmwave_chan,beamspace,sounding} and were extended to utilize the temporal channel correlation in \cite{sounding_beam,tera,madhow}. 
 In \cite{track_ml,track_ml2}, channel tracking was formulated as a maximum likelihood estimation problem.
Sensor-aided beam tracking methods have been proposed recently  \cite{track_sensor,track_sensor1,track_sensor2}. These methods attempted to use an inertial measurement unit (IMU) sensor to assist beam alignment and channel tracking in mmWave systems.  
However, modeling different types of data acquired from sensor and communication signals to design beam tracking methods is difficult. Thus, applying traditional model-based approaches for sensor-based beam tracking is a significant challenge.

Most UEs are hand-held devices. The pose (i.e., location and orientation) of UE devices can vary based on the motion of their human users. This results in dynamic and instantaneous channel variations. In practice, this could cause frequent beam tracking failures. This necessitates the execution of  expensive channel acquisition procedures to recover from failures. Thus, beam tracking algorithms should handle dynamically-varying channels to reduce beam tracking failures and consequently save resource overhead. However, the performance of most existing beam tracking algorithms is limited because they rely on somewhat simple prior linear models to describe time-varying channels. In fact, channels often exhibit structured temporal behavior due to the motion of the UE device.  Therefore, channel models that represent such temporal behavior well are required.  

Recently, deep neural networks (DNNs) have received considerable attention owing to their ability to find an abstract  representation of high-dimensional data \cite{deeplearning}. DNNs can model complex non-linear relationships using multiple layers of an artificial neural network. DNNs have achieved state-of-the-art performance in various challenging machine learning tasks. They have been particularly effective for applications in which the existing analytical models cannot adequately describe the  distribution of the data. Thus, a DNN can be a suitable candidate for modeling the temporal behavior of mmWave channels caused by the motion of a UE device. Recently, a DNN has been applied for beam tracking in mmWave systems in \cite{MLAOA,ML_track2,ML_mmwave_BT}.

In this paper,   we propose an enhanced beam tracking method, that models rapidly-varying mmWave channels using DNNs. We employ a long short-term memory (LSTM) architecture to describe the temporal evolution of the AoAs and AoDs, using the information available in a UE device. Specifically, the LSTM predicts the distribution of the AoA and AoD states for the current beam transmission cycle based on the sequence of the previous channel estimates and IMU sensor signals.  This distribution is used for two main beam tracking operations. First, the distribution of the AoAs and AoDs is used to determine the Rx and Tx sounding beams to be used in the current beam transmission cycle. Second, the predicted channel distribution is used as prior information to  update the channel estimate. The proposed LSTM-based prediction model is incorporated into a sequential Bayesian estimation framework, in which the channel information is updated through a {\it prediction update step} and {\it measurement update step} in an alternating manner. Note that the proposed method uses the LSTM-based prediction model to update the channel distribution in the prediction update step. This distribution is then used as the prior channel information in the subsequent measurement update step. 
  We evaluate the performance of the proposed beam tracking method via computer simulation.
 Our results demonstrate that the proposed method achieves  significant performance gains over conventional beam tracking methods under various high-mobility scenarios.

 The contributions of this paper are summarized as follows;
 \begin{itemize}
     \item Our method uses a DNN to enhance the beam tracking performance in mmWave systems. As compared with widely used simple linear models, the DNN model can capture the complex temporal channel behavior caused by the motion of a device, thereby offering an enhanced beam tracking performance.
      The superiority of the proposed DNN-based beam tracking scheme over the existing methods is confirmed via numerical evaluation. 
      \item We incorporate our DNN-based channel model into the sequential Bayesian filtering framework.  The role of machine learning models is restricted to modeling the temporal behavior of channels only and we use the analytical model describing the relation from the transmitted beam to the   measurements in the measurement update step. This is consistent with the design principles of respecting established models for well-known physical processes and using data-driven approaches only where the actual physical process is barely known (e.g., temporal channel evolution under mobility environment).     This approach contrasts with the end-to-end modeling of beam tracking proposed in \cite{MLAOA}.
           \item Recently, a DNN-based channel tracking has been proposed in \cite{MLAOA}. The method in \cite{MLAOA} directly estimates the AoA using the DNN, whereas the proposed method predicts the future distribution of AoA. The predicted AoA information is then used to update the channel estimate based on the measurement model. Another key difference from the aforementioned method is that the proposed method  utilizes  various types of signals acquired by motion sensors for beam tracking. 
    \end{itemize}

\section{mmWave Beam Tracking Systems} \label{sec:sys_model}    
%\section{MmWave Channel Model and Conventional Beam Training} \label{sec:sys_model}
%In this section, we provide several background information for this paper. 
In this section,  we describe the mmWave channel model and introduce the widely used beam tracking protocol. 
\subsection{MmWave Channel Model} \label{subsec:channel_model}
%To describe the channel model, we consider the downlink channel from the base-station to the  $p$th user.
Recall that the BS and UE have antenna arrays of sizes $N_{b}$ and $N_{m}$, respectively. The downlink channel  from the BS to the UE  can be expressed as the matrix $\mathbf{H}_{t}$ of size $N_{m} \times N_{b}$, where the $(i,j)$th element of $\mathbf{H}_{t}$  represents the channel gain from the $j$th antenna of the BS to the $i$th antenna of the UE.  The subscript $t$ represents the $t$th beam transmission period. The channel  $\mathbf{H}_{t}$ is assumed to be constant within the $t$th beam training period.  The mmWave channel can be represented in the angular domain as  \cite{mmwave_chan,beamspace}    
\begin{align}  
  \mathbf{H}_{t} & = %\sqrt{N_{b}N_{m}}\sum\limits_{l=1}^L\alpha_{p,l,t}\mathbf{a}^{(m)}(\theta^{(m)}_{p,l,t})\left(\mathbf{a}^{(b)}(\theta^{(b)}_{p,l,t})\right)^{H}
  \sum\limits_{l=1}^L\alpha_{l,t}\mathbf{a}^{(m)}(\theta^{(m)}_{l,t})\left(\mathbf{a}^{(b)}(\theta^{(b)}_{l,t})\right)^{H},  \label{eq:physical_channel}
  %\\
%  & = \mathbf{A}_{p,t,n}^{(m)} \mathbf{\Lambda}_{p,t,n} (\mathbf{A}_{p,t,n}^{(b)})^{H},
\end{align}
 where $L$ is the total number of paths, $\alpha_{l,t}$ is the channel gain for the $l$th path, and $\theta^{(b)}_{l,t}$ and $\theta^{(m)}_{l,t}$ are the AoD and AoA, respectively.  The AoD and AoA are obtained from $\theta^{(b)}_{l,t} = \sin(\phi_{l,t}^{(b)})$, $\theta^{(m)}_{l,t} = \sin(\phi_{l,t}^{(m)})$,   
where $\phi_{l,t}^{(b)}$ and $\phi_{l,t}^{(m)}\in[-\frac{\pi}{2},\frac{\pi}{2}]$ are the AoD and AoA in radians, respectively. The steering vectors $ \mathbf{a}^{(b)}(\theta) $ and $\mathbf{a}^{(m)}(\theta)$  are expressed as
\begin{align}
 \mathbf{a}^{(b)}(\theta) & = \frac{1}{\sqrt{N_{b}}}\left[1,e^{\frac{j2\pi d_b \theta}{\lambda}},e^{\frac{j2\pi2d_b\theta}{\lambda}},\cdots,e^{\frac{j2\pi (N_{b}-1)d_b \theta}{\lambda}}\right]^{T} \nonumber \\  
  \mathbf{a}^{(m)}(\theta) & = \frac{1}{\sqrt{N_{m}}}\left[1,e^{\frac{j2\pi d_m \theta}{\lambda}},e^{\frac{j2\pi2d_m\theta}{\lambda}},\cdots,e^{\frac{j2\pi (N_{m}-1)d_m\theta}{\lambda}}\right]^{T}, \nonumber
\end{align}  
where $d_b$ and $d_m$ are the distances between adjacent antennas for the BS and UE, respectively and $\lambda$ is the signal wavelength. In practical scenarios, $L$ tends to be small because only a few paths exhibit dominant energy. Note that the mmWave channel is determined by the set of parameters  ${\mathbf{\gamma}}_{t} = [\alpha_{1,t},\theta^{(m)}_{1,t},\theta^{(b)}_{1,t},...,\alpha_{L,t},\theta^{(m)}_{L,t},\theta^{(b)}_{L,t}]^{T}$.

\subsection{Beam Tracking Protocol} \label{subsec:BT_and_channel_est}
Fig. \ref{fig:beamtracking_protocol} illustrates a typical beam tracking protocol. Without  prior knowledge about the channel, the initial channel acquisition is performed using  the Tx and Rx sounding beams, whose directions are distributed over a wide range. The beam tracking mode starts once the initial channel acquisition is completed. At the $t$th beam transmission period, the beam tracking method uses the channel knowledge to transmit the pilot symbols using significantly fewer sounding beams directed at certain desired directions. 
After beam transmission, the UE updates the channel estimate based on the measurements. These channel estimates are fed back to the BS through a feedback channel or used for data demodulation. 
This beam tracking procedure is repeated in each beam transmission cycle. 
A similar protocol is observed in the 5G standard, where  the SS-burst slot and CSI-RS slot are reserved for the initial channel acquisition and beam tracking, respectively \cite{3gpp.38.211}. 

\subsection{mmWave Channel Estimation}

At the $t$th beam transmission, the BS transmits  $M_b \cdot M_m$ pilot symbols to the UE using $M_b$ Tx beams and $M_m$ Rx beams.  
Let  $\mathbf{f}_{t,1},..., \mathbf{f}_{t,M_b}$ represent the beamforming vectors used for the Tx beams and $\mathbf{w}_{t,1},..., \mathbf{w}_{t,M_m}$ represent the combining vectors for the Rx beams. When an analog beamformer is used, the beamforming and combining vectors are expressed as $\mathbf{f}_{t,i} = \mathbf{a}^{(b)}(\mu_{t,i}^{(b)})$ and $\mathbf{w}_{t,j} = \mathbf{a}^{(m)}(\mu_{t,j}^{(m)})$, respectively, where $\mu_{t,i}^{(b)}$ and $\mu_{t,j}^{(m)}$ are the corresponding directions of the sounding beams. The  vector received in the $t$th beam transmission cycle is expressed as
\begin{align}
y_{t,(i-1)M_m+j} = \mathbf{w}_{t,j}^{H} \mathbf{H}_{t} \mathbf{f}_{t,i} s_{t,i} + n_{t,(i-1)M_m+j}, 
\label{eq:rn}
\end{align}
for $1\leq i \leq M_b$ and $1\leq j \leq M_m$, where $s_{t,i}$ is the pilot symbol  and $n_{t,(i-1)M_m+j}$ is the additive noise. 
Without losing generality, we let $s_{t,i}=1$ in the sequel.
Note that, for each Tx sounding beam,  $M_m$ Rx sounding beams are swept, resulting in 
$M_m \cdot M_u$ transmissions. Combining the received signals in a vector $\mathbf{y}_t$ as  $\mathbf{y}_t = [y_{t,1},...,y_{t,M_b M_m}]^{T}$ and 
using the angular channel representation in (\ref{eq:physical_channel}), we obtain
\begin{align} \label{eq:capR}
\mathbf{y}_t
&= {\rm vec} (\mathbf{W}_t^H \mathbf{H}_t \mathbf{F}_t) + \mathbf{n}_t, \\
&= {\rm vec} \left( \sum\limits_{l=1}^L\alpha_{l,t}\mathbf{W}_t^H \mathbf{a}^{(m)}\left(\theta^{(m)}_{l,t}\right)\left(\mathbf{a}^{(b)}(\theta^{(b)}_{l,t})\right)^{H}  \mathbf{F}_t\right) + \mathbf{n}_t,
\end{align}
where $\text{vec}(\cdot)$ is the vectorization operation\footnote{For example, ${\rm vec}\left(\begin{bmatrix} 1 & 2 \\ 3 & 4 \end{bmatrix}\right) = [1,3,2,4]^{T}$.}, $\mathbf{n}_t = [n_{t,1},...,n_{t,M_b M_m}]^{T}$, $\mathbf{W}_t = \begin{bmatrix}\mathbf{w}_{t,1} & ... & \mathbf{w}_{t,M_m} \end{bmatrix}$, and $\mathbf{F}_t = \begin{bmatrix}\mathbf{f}_{t,1} & ... & \mathbf{f}_{t,M_b}\end{bmatrix}$. We assume that the channel gains $\alpha_{1,t},...,\alpha_{L,t}$ vary slowly so that they can be assumed as being estimated accurately. 
For specified  $\mathbf{F}_t$ and $\mathbf{W}_t$, the channel estimation problem is equivalent to the determination of  the set of parameters  $\gamma_{t} =[\gamma_{1,t}^{T},...,\gamma_{L,t}^{T}]^{T} =  [[\theta^{(m)}_{1,t},\theta^{(b)}_{1,t}],..., [\theta^{(m)}_{L,t},\theta^{(b)}_{L,t}]]^{T}$.
Accordingly, we formulate  the following state-space equation:
\begin{itemize}
\item State evolution model  
\begin{align}
    \gamma_t &= \mathbf{A}_t \gamma_{t-1} + \mathbf{v}_t, \label{eq:sevol} 
 \end{align}
 %\textcolor{blue}{where the state transition matrix $\mathbf{A}_t = \text{diag} ([\psi^{(m)}_{1,t},\psi^{(b)}_{1,t}, \cdots ,\psi^{(m)}_{L,t},\psi^{(b)}_{L,t}])$ and $\mathbf{v}_t$ is a complex Gaussian vector $CN(0,V_t)$.}
 where $\mathbf{A}_t$ is the auto-regressive parameter and $\mathbf{v}_t$ is a complex Gaussian vector $CN(0,V_t)$.
\item Measurement model
\begin{align}
       \mathbf{y}_t  = {\rm vec} \left( \sum\limits_{l=1}^L\alpha_{l,t}\mathbf{W}_t^H \mathbf{a}^{(m)}\left(\theta^{(m)}_{l,t}\right)\left(\mathbf{a}^{(b)}(\theta^{(b)}_{l,t})\right)^{H}  \mathbf{F}_t\right) + \mathbf{n}_t. \label{eq:meas_evol}
\end{align}
\end{itemize}
Owing to the nonlinearity of the state-space equation, we can use nonlinear Bayesian filtering algorithms. A popular method used in this regard is the extended Kalman filter (EKF) 
\begin{enumerate}
\item Prediction update step
\begin{align} 
      \hat{\gamma}_{t|t-1} &= \mathbf{A}_t \hat{\gamma}_{t-1|t-1}  \nonumber \\
      \mathbf{P}_{t|t-1} &= \mathbf{A}_t \mathbf{P}_{t-1|t-1} \mathbf{A}_t^{H} + \mathbf{V}_t, \label{eq:ekf_pred2}
\end{align}
\item Measurement update step
\begin{align} \label{eq:ekf_mus}
       &\mathbf{K}_t = \mathbf{P}_{t|t-1}\mathbf{O}_{t}^H
    \left(\mathbf{O}_{t}\mathbf{P}_{t|t-1}\mathbf{O}_{t}^H+\sigma_t^2I\right)^{-1} \nonumber \\
    &\mathbf{P}_{t|t} = (I-\mathbf{K}_t\mathbf{O}_{t})\mathbf{P}_{t|t-1} \nonumber \\
    &{\widehat {\mathbf{\gamma}}}_{t|t}={\widehat {\mathbf {\mathbf{\gamma}}}}_{t|t-1}+{\mathbf{K}}_{t}\left(\mathbf{y}_t - 
      q( \widehat {\mathbf {\mathbf{\gamma}}}_{t|t-1})\right),
\end{align}
\end{enumerate}
where the vector $q({\mathbf {\mathbf{\gamma}}}_{t})$ and 
Jacobian matrix $\mathbf{O}_{t}$  are expressed as
\begin{align*}
    q({{\mathbf{\gamma}}}_{t}) & = \sum\limits_{l=1}^L {\rm vec} \left(\alpha_{l,t}\mathbf{W}_t^H \mathbf{a}^{(m)}\left(\theta^{(m)}_{l,t}\right)\left(\mathbf{a}^{(b)}(\theta^{(b)}_{l,t})\right)^{H}  \mathbf{F}_t\right) \nonumber \\
    \mathbf{O}_{t} & = \left.\frac{\partial {q}({\mathbf{\gamma}}_{t})}{\partial\mathbf{\gamma}_{t}}     \right|_{\mathbf{\gamma}_{t}=\mathbf{\hat{\gamma}}_{t|t-1}}.
\end{align*}
The expression for  $\mathbf{O}_{t}$ is provided in Appendix A.
As the prior channel model in (\ref{eq:sevol}) captures only the first-order dynamics of channel variations, EKF often fails to track the complex channel dynamics in the prediction update step, resulting in a large linearization error in the measurement update step. %Furthermore, 

\section{Review of LSTM Model}      

	\begin{figure*}[t]
    \centering
%\includegrachics{beamtrackingprotocolfinal.eps}
    \includegraphics[width=110mm]{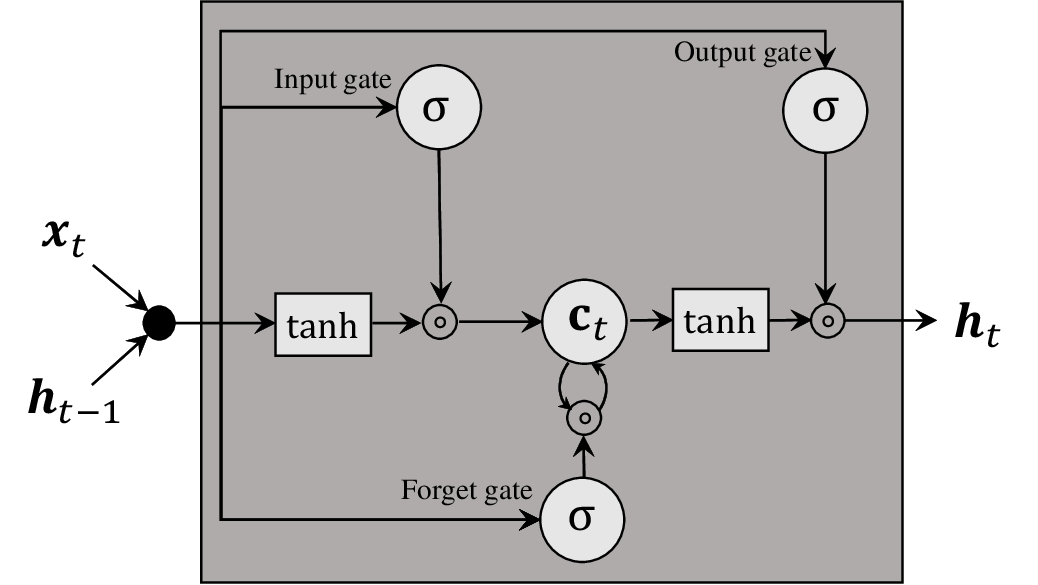}
    \caption{Structure of the basic LSTM model}
    \label{fig:lstm}
\end{figure*}
	
The LSTM is a DNN architecture widely used to analyze time-series data.  Fig. \ref{fig:lstm} depicts the structure of the LSTM. The LSTM uses  recurrent connections to extract  features from  sequence data and stores them in a memory called {\it cell state}. When unfolded in time, the connection from the input to the output in the LSTM is deep in time. This enables an efficient representation of long sequences. The LSTM has been successfully applied to various machine-learning problems, e.g., natural language processing, speech recognition, and machine translation.  The LSTM consists of a cell state, and input, output, and forget gates. The input, output, and forget gating functions can control the information flows entering and leaving the cell state. These gating functions are designed to address the {\it vanishing gradient problems}, in which the gradient signals attenuate considerably in learning long-term dependency \cite{lstm}. 
Whenever the input $x_t$ is fed into the LSTM, the cell state $c_t$ at the time step $t$ is
updated according to the following recursive equations
\begin{align}
	i_t &= \sigma(W_{xi} x_t + W_{hi}h_{t-1} + b_i)   \\
	f_t &= \sigma(W_{xf} x_t + W_{hf}h_{t-1} + b_f)   \\
	o_t &= \sigma(W_{xo} x_t + W_{ho}h_{t-1} + b_o)  \\
	g_t &= \tanh(W_{xc} x_t + W_{hc}h_{t-1} + b_c)   \\
	c_t &= f_t \odot c_{t-1} + i_t \odot g_t \\
	h_t &= o_t \odot \tanh(c_t),
\end{align}
	where  
\begin{itemize}
	\item $\sigma(x) = \frac{1}{1+e^{-x}}$: sigmoid function
	\item $a \odot b$: element-wise product
    \item $W_{xi}, W_{hi}, W_{xf}, W_{hf}, W_{xo},     W_{ho},W_{xc},W_{hc}$: weight matrices for linear transformation
	\item $b_i, b_f, b_o, b_c$: bias vector
	\item $i_t$: input gating vector
	\item $f_t$: forget gating vector
	\item $o_t$: output gating vector
	\item $g_t$: state update vector
	\item $h_t$:  output hidden state vector.
\end{itemize}
The output $h_t$ of the LSTM contains the feature required to perform the specified task. The desired output can be determined from the feature $h_t$ through an additional neural network. The LSTM is trained to minimize the appropriately designed loss function using the back-propagation through time (BPTT) algorithm.  

\section{Proposed Deep Learning-based Beam Tracking} \label{sec:proposed}

\begin{figure*}[t]
    \centering
%\includegrachics{beamtrackingprotocolfinal.eps}
    %\includegraphics[width=\textwidth]{figures/Overall_system.eps}
    %\includepdf[width=\textwidth]{figures/overall.pdf}
        \includegraphics[width=\textwidth]{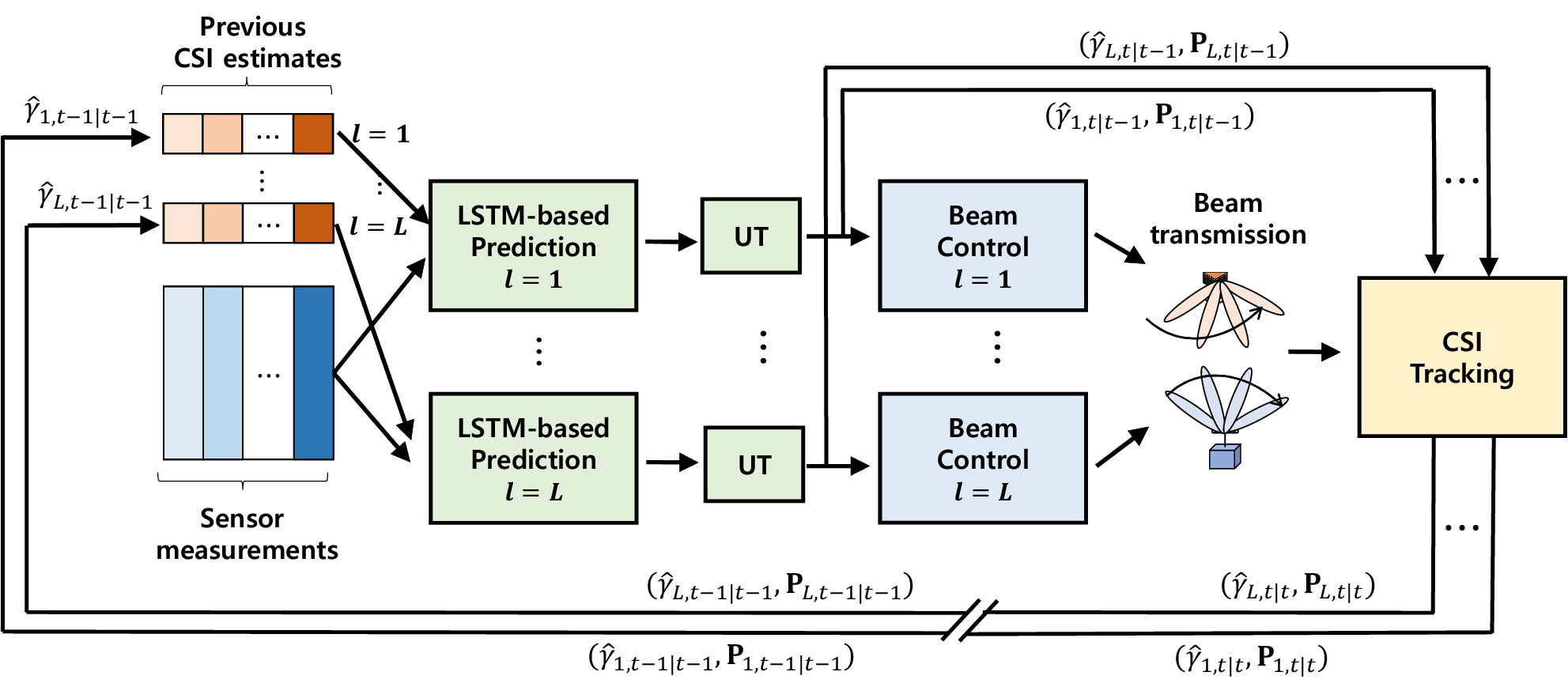}
    \caption{Block diagram of the proposed beam tracking method}
    \label{fig:system_overview}
\end{figure*}

In this section, we describe the proposed beam tracking method. 
The structure of the overall system is depicted in Fig. \ref{fig:system_overview}. The LSTM-based prediction model predicts the distribution of the channel state at the $t$th beam training period based on all the previously available channel estimates and IMU sensor signals. The prediction model produces the mean and covariance matrix of AoD and AoA separately for each path. Note that the parameters of each model  are shared among $L$ paths. As shown in Fig. \ref{fig:system_overview}, the output of the prediction model is used to steer both Tx and Rx sounding beams and update the channel estimates based on the received beams.

\subsection{LSTM-based Channel Prediction}

\begin{figure}[t]
    \centering
%\includegrachics{figures/beamtrackingprotocolfinal.eps}
    \includegraphics[width=120mm]{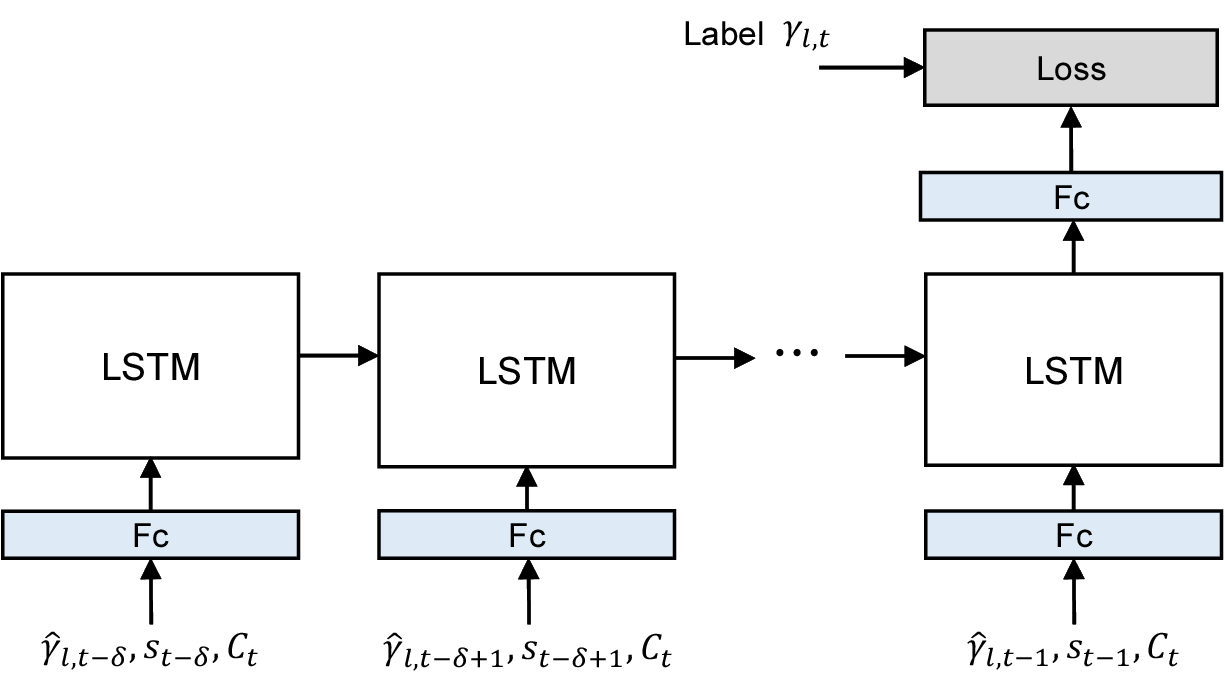}
    \caption{Structure of the LSTM-based prediction model}
    \label{fig:architecture}
\end{figure}

The input to the LSTM-based prediction model includes
\begin{itemize}    
  \item $\hat{\gamma}_{l,t-\delta:t-1}=\left[\hat{\gamma}_{l,t-\delta}^{T},...,\hat{\gamma}_{l,t-1}^{T}\right]^{T}$: the sequence of the previous $\delta$ channel estimates acquired before the $t$th beam transmission period begins.
  \item $s_{t-\delta:t-1}=[s_{t-\delta}^T,...,s_{t-1}^T]^{T}$: the sequence of the previous sensor signal samples of $J$ types. For example,  with $J=3$, the vector $s_{t-i}$ contains the velocity, angular velocity and angular acceleration samples acquired from the IMU sensor. As the sampling frequency of these signals can be different from that of the beam transmissions, the sensor signals can be resampled to produce $K$ samples for each beam transmission cycle. Finally, the vector $s_{t-i}$ is filled with $KJ$ signal samples.
   \item $\mathcal{C}_{t}$: the vector that represents contextual information such as the location and activity of the UE. Although it does not represent sequential data,  contextual information provides supplementary information on the channels.
 \end{itemize} 
The proposed prediction model aims to determine the distribution $p({\gamma}_{l,t}|\hat{\gamma}_{l,t-1:t-\delta},s_{t-1:t-\delta},\mathcal{C}_{t})$ for each channel path for the given past channel estimates $\hat{\gamma}_{l,t-\delta:t-1}$, sensor signals $s_{t-\delta:t-1}$, and context information $\mathcal{C}_{t}$.   We employ the LSTM  to model the dependencies  between the input and the future channel state. The structure of the LSTM-based prediction model is depicted in Fig.~\ref{fig:architecture}. The signal samples $\{\hat{\gamma}_{l,t-\delta}, s_{t-\delta}, \mathcal{C}_t\}$, ..., $\{\hat{\gamma}_{l,t-1}, s_{t-1}, \mathcal{C}_t\}$ are encoded separately by the {\it input fully-connected (Fc) layers}, i.e., 
\begin{align}
\nu_{l,t-i} = {\rm Fc}(\{\hat{\gamma}_{l,t-i}, s_{t-i}, \mathcal{C}_t\}),
\end{align}
where $\nu_{l,t-i}$ is the embedding vector obtained by the Fc layers.  The embedding vectors are fed to the LSTM one by one to update the cell state. After $\delta$  update of the LSTM, the output $h_t$ is fed into the {\it output Fc layers} to produce the estimate $\hat{\gamma}_{l,t}$. That is,   \begin{align}
\hat{\gamma}_{l,t} = {\rm Fc} \left({\rm LSTM} \left(\{ \nu_{l,t-\delta},...,\nu_{l,t-1} \} \right)\right).
\end{align}

The parameters of the LSTM and Fc layers are determined in the training procedure. In practice, the training data could be collected by deploying several reference UEs, that log the channel states and sensor signals in real scenarios. The model is trained to minimize the negative log-likelihood loss $ \sum \|{\gamma}_{l,t} - \hat{\gamma}_{l,t} \|^2 $, where 
  the ground truth ${\gamma}_{l,t}$ can be obtained directly from the training data. 
The prediction network is trained using the standard BPTT algorithm with the ADAM optimization \cite{adam}. The model weights are updated over a minibatch of size {\it MINIBATCH}. The training starts with the initial learning rate {\it LEARNING\_RATE}. The learning rate decays by half in each {\it DECAY\_EPOCH} epochs. Note that the model is trained over a total of {\it TOTAL\_EPOCH} epochs.

\subsection{Channel Tracking}  

The sequential Bayesian estimation framework is widely used to estimate time-varying channels. The Bayesian principle involves  updating the distribution of a channel based on all the information available at each step. By adopting this principle, we update the mean ${\widehat {\mathbf{\gamma}}}_{l,t-1|t-1}$ and the covariance matrix $\mathbf{P}_{l,t-1|t-1}$ by ${\widehat {\mathbf{\gamma}}}_{l,t|t-1}$ and $\mathbf{P}_{l,t|t-1}$ through the prediction update step. Subsequently, the measurement update step updates ${\widehat {\mathbf{\gamma}}}_{l,t|t-1}$ and $\mathbf{P}_{l,t|t-1}$ by ${\widehat {\mathbf{\gamma}}}_{l,t|t}$ and $\mathbf{P}_{l,t|t}$. In the prediction update step,  the LSTM-based prediction model is used to obtain ${\widehat {\mathbf{\gamma}}}_{l,t|t-1}$ and $\mathbf{P}_{l,t|t-1}$. As the LSTM-based prediction model produces the point estimate of the future channel state, 
 we employ the unscented transformation (UT) \cite{ut,ukf} to obtain the distribution. First, for given ${\widehat {\mathbf{\gamma}}}_{l,t-1|t-1}$ and  $\mathbf{P}_{l,t-1|t-1}$, we generate  $2P+1$ sigma vectors $\chi_i$ with the corresponding weights $w_i$, i.e.,  
\begin{align}
    \chi_0 &= {\widehat {\mathbf{\gamma}}}_{l,t-1|t-1} \\
    \chi_i & = {\widehat {\mathbf{\gamma}}}_{l,t-1|t-1} + \left(\sqrt{(L+\lambda)\mathbf{P}_{l,t-1|t-1}} \right)_i   \;\;\;\;\; i=1,...,P \\
      \chi_i & = {\widehat {\mathbf{\gamma}}}_{l,t-1|t-1} - \left(\sqrt{(L+\lambda)\mathbf{P}_{l,t-1|t-1}} \right)_{i-P}   \;\;\;\;\; i=P+1,...,2P \\
      w_0^{(m)} & = \lambda/(L+\lambda) \\
      w_0^{(c)} & = \lambda/(L+\lambda) + (1-\alpha^2 +\beta) \\
      w_i^{(m)} & = w_i^{(c)} = 1/(2(K+\lambda)),
\end{align}
where $\lambda = \alpha^2 P-P$ is a scaling parameter, $\alpha=10^{-3}$ determines the spread of the sigma points around ${\widehat {\mathbf{\gamma}}}_{t-1|t-1}$, and $\beta=2$ reflects the prior distribution of ${\gamma}_{l,t}$. $\left(\sqrt{(L+\lambda)\mathbf{P}_{t-1|t-1}} \right)_i$  is the $i$th row of the matrix square root. These sigma vectors are passed through the LSTM-based prediction model, which generates $2P+1$ outputs $\nu_0,...,\nu_{2P}$. Finally, the updated distribution ${\widehat {\mathbf{\gamma}}}_{l,t|t-1}$ and $\mathbf{P}_{l,t|t-1}$ are obtained by the weighted sums
\begin{align}
    {\widehat {\mathbf{\gamma}}}_{l,t|t-1} &= \sum_{i=0}^{2P} w_i^{(m)}\nu_i \label{eq:ws1} \\
    \mathbf{P}_{l,t|t-1} &= \sum_{i=0}^{2P} w_i^{(c)} (\nu_i - {\widehat {\mathbf{\gamma}}}_{l,t|t-1} )(\nu_i-{\widehat {\mathbf{\gamma}}}_{l,t|t-1} )^T. \label{eq:ws2}
    \end{align}
It was shown in \cite{ut} that  the UT yields  approximations that are accurate up to at least the second order, with the accuracy
of the third and higher-order moments determined by the selection
of $\alpha$ and $\beta$.
After the prediction update step is completed, the measurement update step is performed to obtain the statistics ${\widehat {\mathbf{\gamma}}}_{l,t|t}$ and $\mathbf{P}_{l,t|t}$ from 
\begin{align} 
       &\mathbf{K}_t = \mathbf{P}_{t|t-1}\mathbf{O}_{t}^H
    \left(\mathbf{O}_{t}\mathbf{P}_{l,t|t-1}\mathbf{O}_{t}^H+\sigma_t^2I\right)^{-1} \nonumber \\
    &\mathbf{P}_{l,t|t} = (I-\mathbf{K}_t\mathbf{O}_{t})\mathbf{P}_{l,t|t-1} \nonumber \\
    &{\widehat {\mathbf{\gamma}}}_{l,t|t}={\widehat {\mathbf {\mathbf{\gamma}}}}_{l,t|t-1}+{\mathbf{K}}_{t}\left(\mathbf{y}_t - 
      q( \widehat {\mathbf {\mathbf{\gamma}}}_{l,t|t-1})\right). \label{eq:meas222}
\end{align} 
 Note that  this measurement update step  is equivalent to that of the EKF.  

\subsection{Predictive Beam Control}
The direction of the sounding beams needs to be determined based on the best available channel information in the $t$th beam transmission cycle.   Before the sounding beams are transmitted, the best available channel information is the statistics ${\widehat {\mathbf{\gamma}}}_{l,t|t-1}$ and $\mathbf{P}_{l,t|t-1}$ obtained using the UT. 
Based on the Gaussian approximation  $p({\gamma}_{l,t}|\hat{\gamma}_{l,t-1:t-\delta},s_{t-1:t-\delta},\mathcal{C}_{t}) \approx N({\widehat {\mathbf{\gamma}}}_{l,t|t-1},\mathbf{P}_{l,t|t-1})$, we determine the beam angles $\mu_{t,1}^{(b)},...,\mu_{t,M_b}^{(b)}$ and $\mu_{t,1}^{(m)},...,\mu_{t,M_m}^{(m)}$, where $\mathbf{f}_{t,i} = \mathbf{a}^{(b)}(\mu_{t,i}^{(b)})$ and $\mathbf{w}_{t,j} = \mathbf{a}^{(m)}(\mu_{t,j}^{(m)})$. 
The optimal beam angles can be determined by maximizing the expected channel estimation performance with respect to the parameters $\mu_{t,1}^{(b)},...,\mu_{t,M_b}^{(b)}$ and $\mu_{t,1}^{(m)},...,\mu_{t,M_m}^{(m)}$. An in-depth study on the optimization of the sounding beams has been presented in \cite{Dedicated,pomdp,mmwave_chan}. In our previous work \cite{Dedicated}, the problem of sounding beam adaptation was formulated as a minimization of the Cramer-Rao lower bound (CRLB) of the channel estimation error over the combinations of beam codebook indices.  When only two sounding beams are used for Tx and Rx, i.e., $M_b=M_m=2$, the optimal beam directions could be determined by a two-dimensional search over the beam codebook. In this study, adopting the method in \cite{Dedicated}, the CRLB is derived for the given channel distribution $N({\widehat {\mathbf{\gamma}}}_{l,t|t-1},\mathbf{P}_{l,t|t-1})$, and   the optimal sounding beam angles are determined.  With the setup $M_b=M_m=2$, we choose the values of the beam angles $\mu_{t,1}^{(b)},\mu_{t,2}^{(b)},\mu_{t,1}^{(m)},\mu_{t,2}^{(m)}$ using two-dimensional search.  Note that this beam control algorithm  moves the group of  sounding beams  toward the future AoD and AoA directions in advance.

\subsection{Algorithm Summary} 
The proposed beam tracking algorithm is summarized in Algorithm \ref{alg:proposed}. 

\begin{algorithm} [t] 
	\caption{Proposed beam tracking algorithm}
	\begin{algorithmic}[1]
	\State At the $t$th beam transmission cycle
	\State Input: ${\widehat {\mathbf{\gamma}}}_{1,t-1|t-1},...,{\widehat {\mathbf{\gamma}}}_{L,t-1|t-1}$ and $\mathbf{P}_{1,t-1|t-1},...,\mathbf{P}_{L,t-1|t-1}$
	\State  {\bf Prediction update step:}
	\For {$i=1$ to L $...$}
	\State Generate $2P+1$ sigma samples $\chi_0,...,\chi_{2P}$ according to ${\widehat {\mathbf{\gamma}}}_{l,t-1|t-1}$ and  $\mathbf{P}_{l,t-1|t-1}$. 
	\State 	  Generate $2P+1$ output samples $\nu_0,...,\nu_{2P}$ using the LSTM prediction model.  
	\State Update ${\widehat {\mathbf{\gamma}}}_{l,t|t-1}$ and  $\mathbf{P}_{l,t|t-1}$ according to (\ref{eq:ws1}) and (\ref{eq:ws2}).
		\EndFor
	\State  {\bf Beam adaptation and transmission:}
		\State Determine the directions of the sounding beams using the method in \cite{Dedicated} and transmit the beams accordingly.
		\State  {\bf Measurement update step:}
		\For {$i=1$ to L $...$}
	%\State {\bf Beam transmission is performed}
			\State Update ${\widehat {\mathbf{\gamma}}}_{l,t|t}$ and  $\mathbf{P}_{l,t|t}$ according to (\ref{eq:meas222}).
		\EndFor
		\State $t \leftarrow t+1$ and go back to line 1.

	\end{algorithmic}
	\label{alg:proposed}
\end{algorithm}

\section{Experimental Results} \label{sec:simul}

In this section,  we evaluate the performance of the proposed beam tracking algorithm. 

\subsection{Simulation Setup}
\subsubsection{MmWave System Setup}

In our simulations, we considered 28GHz frequency band communications with uniform linear array (ULA) antennas whose adjacent elements are spaced by a half wavelength. We considered the communication between a single BS with $N_b=32$ Tx antennas and  a single UE device with $N_m = 32$ Rx antennas. 
%Fig. \ref{fig:frame} shows the frame structure considered in the simulations.
Following the 5G NR standard \cite{3gpp.38.211}, the symbol duration over which a single beam is transmitted was set to $8.93 \mu s$ and 14 symbols are included in each slot of duration $125 \mu s$. 
In the simulations, $4$ sounding beams ($M_b=2$ and $M_m=2$) were transmitted every $T_{CSI}$ slots. The periodicity of beam transmission $T_{CSI}$ was set to 160 slots based on the 5G NR standard \cite{3gpp.38.211}. We use the beam codebook,   which contains 64 beams with equally spaced angles. 
The symbol slots not used for beam tracking were allocated for data transmission.  
The data symbols in each slot were modulated using binary phase-shift keying (BPSK) modulation. The data precoding and combining matrices were obtained from the left and right singular vectors of the channel matrix associated with the highest singular value, respectively. 

 \subsubsection{Mobility Model}

\begin{figure}[!t]
    \centering
    \subfigure[]{
    \includegraphics[width=90mm]{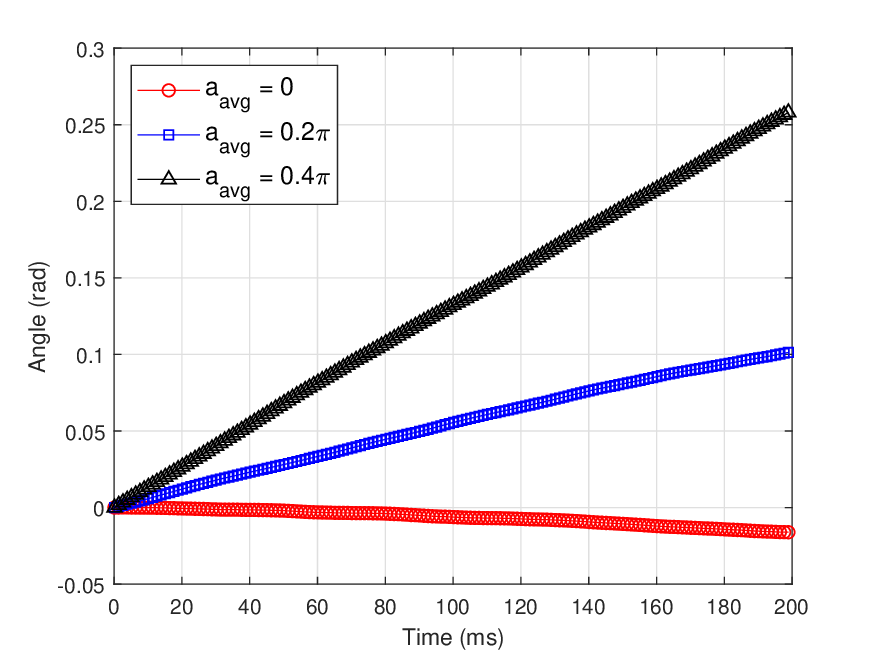}   
    }
    \subfigure[]{
    \includegraphics[width=90mm]{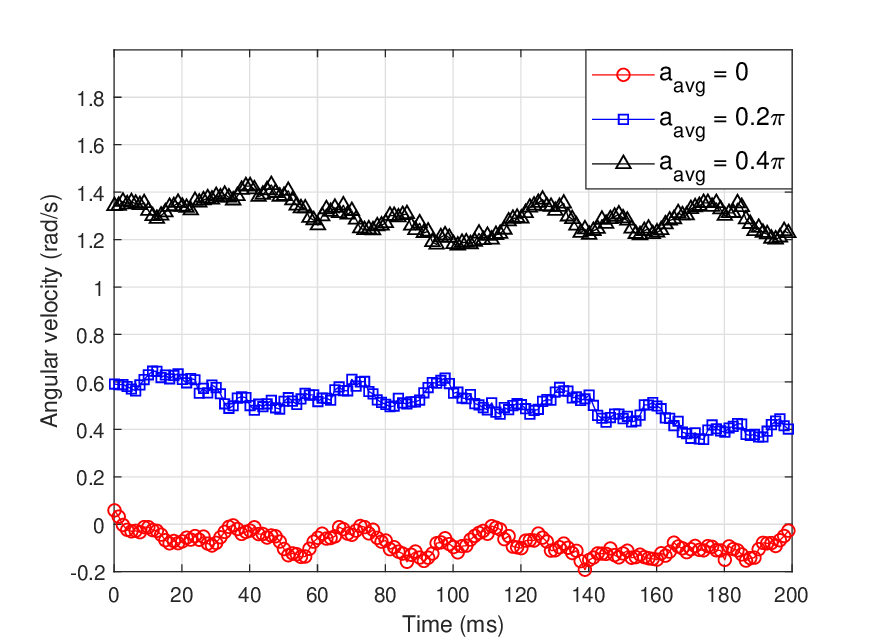}
    }
    \caption{Variation of (a) AoA and (b) angular velocity for several values of the parameter $a_{avg}$.}
    \label{fig:angle_variation}
\end{figure}

We assume that  the BS is located sufficiently far away from the UE and the AoD does not change considerably in time. On the contrary, owing to the motion of the UE, the AoA varies considerably. Thus, we assume that only AoA varies in time according to the following dynamic model:  
\begin{align} 
    a_{l,n} &= (1-\rho) a_{avg} + \rho a_{l,n-1} + w_t  \\
    \theta_{l,n}^{(m)} &= \theta_{l,n-1}^{(m)} +  \Delta t ( a_{l,n} ),  \label{eq:aoagen}
\end{align}   
where $n$ is the slot index, $a_{l,n}$  denotes the angular velocity of AoA for the $l$th path, $\theta_{l,n}^{(m)}$ denotes the AoA, $a_{avg}$ denotes the average velocity, $\Delta t=125us$ is the symbol duration, $\rho=0.9999$ is the auto-regressive (AR) parameter, and  $w_{t} \sim N(0,0.2(1-\rho^2))$. Note that the angular velocity $a_{l,n}$ is modeled by the AR process, and the AoA $\theta_{l,n}^{(m)}$ is generated by accumulating the velocity.  We assume that the channel gain and AoD are constant and known.  
The number of paths $L$ was set to 3. The AoAs were generated independently for each path.
Fig. \ref{fig:angle_variation} shows the change in the AoA and angular velocity with different values of $a_{avg}$. 
A higher value of $a_{avg}$ leads to faster motion and consequently more dynamic AoA variations.  Note that $a_{avg}=0.4\pi$ indicates that the UE device rotates in approximately $2.5 s$.

\subsubsection{LSTM-based Prediction Model}

We present the detailed configurations of the proposed prediction model. The length  of the input sequence for updating the LSTM was set to $\delta = 3$. Each input vector consists of \begin{itemize}
    \item The previous channel estimate
    \item $K=4$ samples of angular velocity sensor measurements ($rad/s$)
    \item $K=4$ samples of angular acceleration sensor measurement ($rad^2/s$).
\end{itemize}
The IMU sensor measurements were generated by computing the first and second sample derivatives of the AoA and adding Gaussian noise. The signal-to-noise ratio (SNR) was set to 5 dB when testing the prediction model. 
We assume that the sampling period of the IMU sensor signals is $K=4$ times lower than $T_{CSI}$. 
The FC layers at the input and  output have 16 and 32 hidden nodes, respectively. The LSTM uses the stacked cell states of size 32.

\subsubsection{Training Procedure}
      
The training configurations  are as follows:     
\begin{itemize}
    \item {\it MINIBATCH : 64} 
    \item {\it INITIAL\_LEARNING\_RATE : 0.01}
    \item {\it DECAY\_EPOCH : 3}  
    \item {\it DECAY\_RATE : 0.1}
    \item {\it TOTAL\_EPOCH : 30}
    \item {\it OPTIMIZER : Adam optimizer}
\end{itemize}
A total of $3,000,000$ data examples were generated for training and 1,000,000 examples were used to evaluate the proposed beam tracking method. The training data were generated with a random SNR uniformly distributed in the range $ [6,15]$ dB. The SNR of the sensor signals was also randomly determined  from the same candidate set.

\subsection{Experimental Results} \label{sec:simul_result}

\begin{figure} [!ht]  
	\centering   
	%\subfigure[Average speed $= 0.1\pi$, $T_{CSI}=80$]{      
    %\includegraphics[width=75mm]{figures/ber_80_1.eps}
    %}
    \subfigure[]{
    \includegraphics[width=83mm]{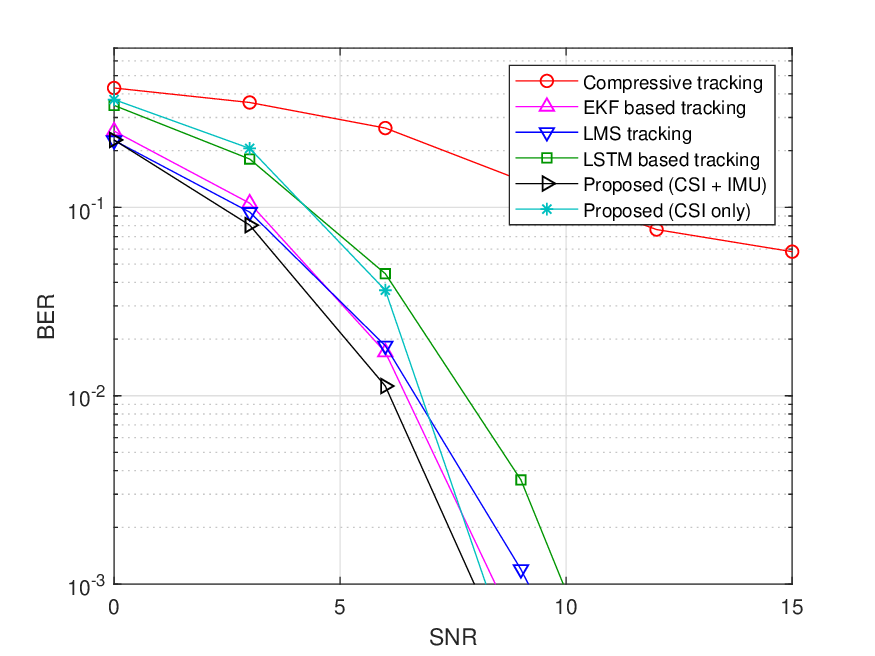}
    }
	%\subfigure[Average speed $= 0.2\pi$, $T_{CSI}=80$]{      
    %\includegraphics[width=75mm]{figures/ber_80_2.eps}
    %}
    \hspace{-1.0cm}
    \subfigure[]{
    \includegraphics[width=83mm]{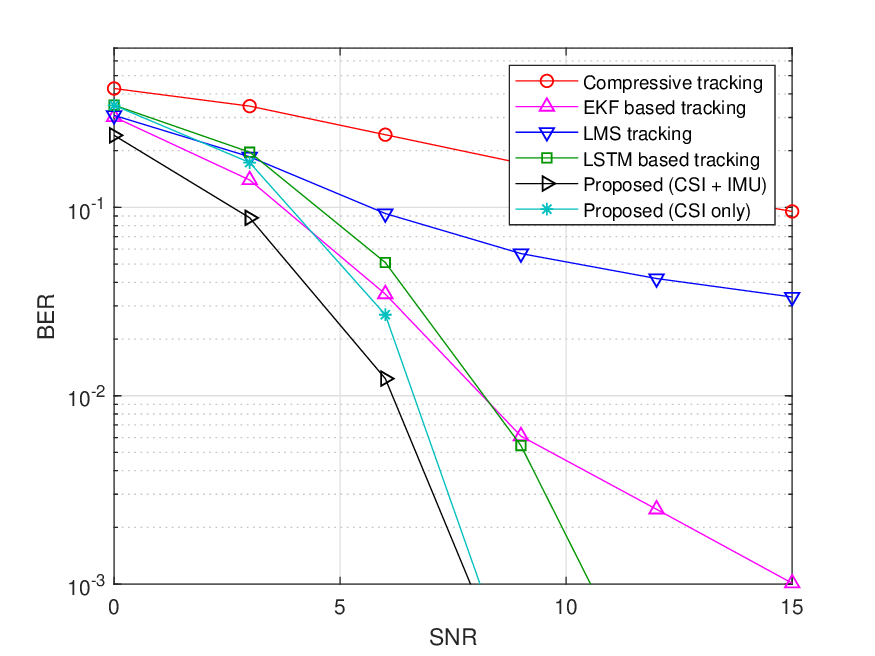}
    }
    %\subfigure[Average speed $= 0.4\pi$, $T_{CSI}=80$]{      
    %\includegraphics[width=75mm]{figures/ber_80_4.eps}
    %
    \subfigure[]{
    \includegraphics[width=83mm]{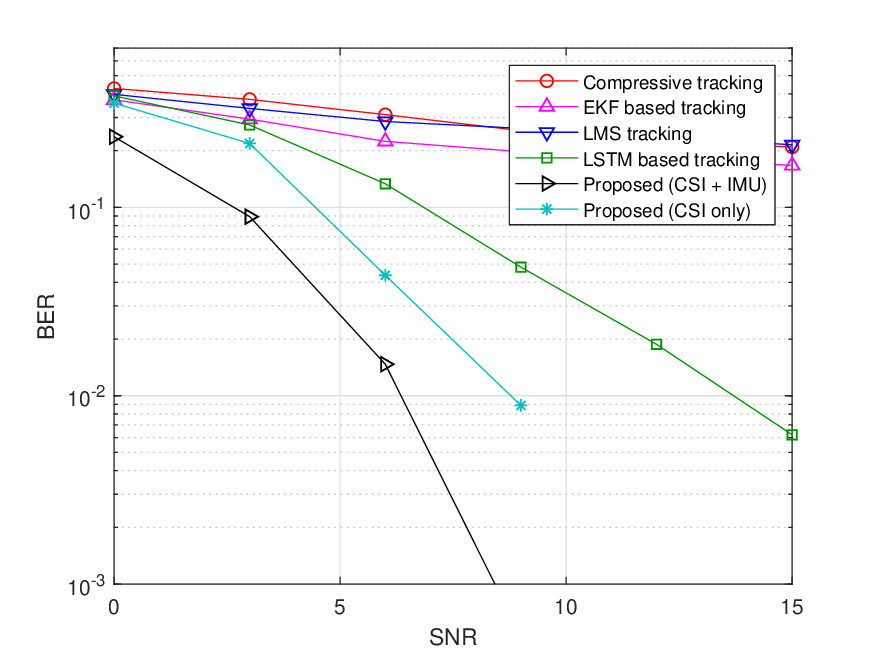}
    }
\caption{BER versus SNR of several channel tracking methods for (a) $a_{avg}=0.1\pi$, (b) $a_{avg}=0.2\pi$, and (c) $a_{avg}=0.4\pi$. }   
\label{fig:ber_SNR}
\end{figure}

\begin{figure} [!t]  
	\centering   
	%\subfigure[Average speed $= 0.1\pi$]{      
    %\includegraphics[width=75mm]{figures/mse_80_1.eps}
    %}
    \subfigure[]{
    \includegraphics[width=83mm]{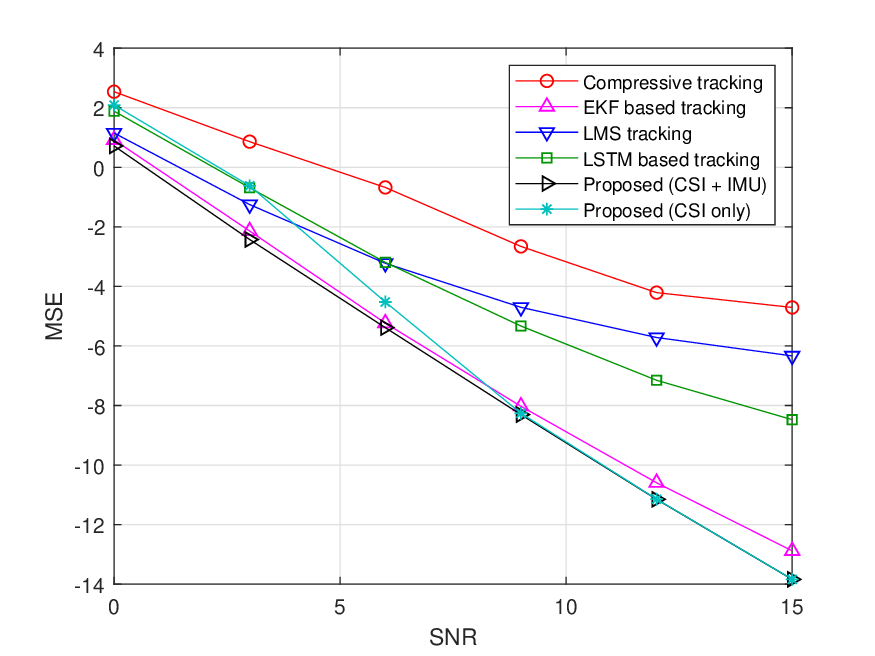}
    }
	%\subfigure[Average speed $= 0.2\pi$]{      
    %\includegraphics[width=75mm]{figures/mse_80_2.eps}
    %}
        \hspace{-1.0cm}
    \subfigure[]{
    \includegraphics[width=83mm]{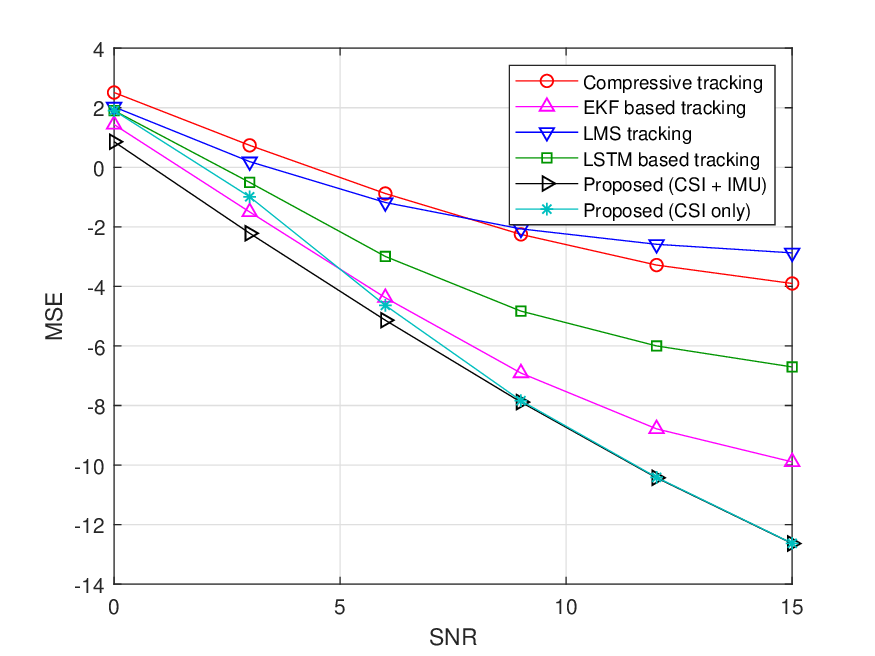}
    }
    %\subfigure[Average speed $= 0.4\pi$, $T_{CSI}=80$]{      
    %\includegraphics[width=75mm]{figures/mse_80_4.eps}
    %}
    \subfigure[]{  
    \includegraphics[width=83mm]{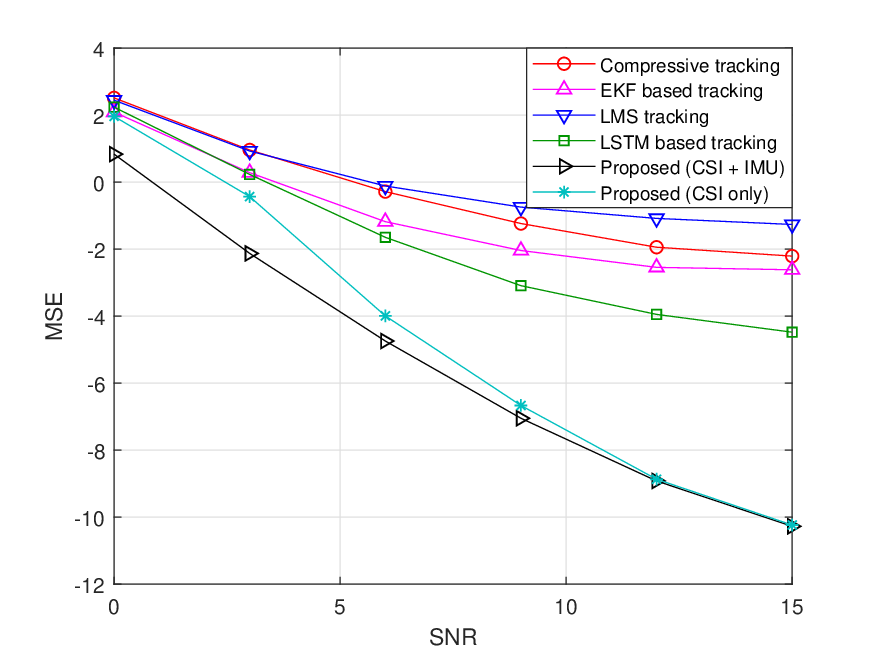}
    }
\caption{Normalized MSE versus SNR of several channel tracking methods for (a) $a_{avg}=0.1\pi$, (b) $a_{avg}=0.2\pi$, and (c) $a_{avg}=0.4\pi$. }   
\label{fig:mse_SNR}
\end{figure}

In this section, we compare our method with the following mmWave channel tracking  methods: 
\begin{enumerate}
\item Compressive channel tracking \cite{madhow}:  Orthogonal matching pursuit \cite{omp} followed by off-grid refinement was used to track the  AoA.  
\item EKF method \cite{ekf}:  The  AoA was estimated using the EKF.   
\item Least mean square (LMS) method  \cite{track_ml}: The AoA was estimated by using the LMS filter. 
\item LSTM based tracking \cite{MLAOA}: The LSTM model directly estimates the AoA. It was trained using the cosine loss function. 
\item Proposed (CSI) method:  Only previous AoA estimates were used to predict the future channel state information (CSI). This method was evaluated to investigate the advantage of using IMU sensors for beam tracking.  
\item Proposed (CSI+IMU) method: The previous AoA estimates and IMU measurements were used to predict the future channel distribution.         
\end{enumerate}  
As the compressive channel tracking, LMS, and LSTM-based tracking methods do not produce the distribution of AoA, we used  $M_m=2$ beams closest to the previous AoA estimate as the Rx sounding beams. In contrast,  like the proposed method, the EKF method yields the distribution of the AoA, which is used to determine the Rx sounding beams. 
The normalized mean square error (MSE) is defined as
\begin{align*}
    MSE = 10 \log_{10} \frac{\left\| {\mathbf{H}_{t}-\mathbf{\hat{H}}_{t}} \right\|_F^2}{\left\|{\mathbf{H}_{t}}\right\|_F^2}.
\end{align*}

Fig.~\ref{fig:ber_SNR} shows the bit error rate (BER) performance as a function of SNR. The parameter $a_{avg}$ indicates the extent of mobility of the UE. Fig.~\ref{fig:ber_SNR} (a), (b), and (c) show the performance curves for $a_{avg}=0.1 \pi$, $0.2\pi$, and $0.4\pi$, respectively.  We observe that the proposed (CSI+IMU) method outperforms the existing methods for all the cases considered.  When $a_{avg}$ is $0.1 \pi$ rad/s, the proposed method achieves a performance gain of approximately 1 dB over the EKF method at the BER of $10^{-3}$. As $a_{avg}$ increases, the channel changes more dynamically and the performance gain of the proposed method increases. With $a_{avg} = 0.2\pi$, the proposed method achieves a gain of more than 3 dB over other algorithms. Furthremore, with $a_{avg} = 0.4\pi$, the performance gain increases up to more than 10 dB. This indicates that the LSTM-based channel model provides a more accurate model of time-varying AoAs, and thus, superior performance is achieved under higher mobility.    Fig.~\ref{fig:ber_SNR} also shows the advantage of using IMU sensors for beam tracking. The proposed (CSI+IMU) method achieves a performance gain over the proposed (CSI) method, especially in the low SNR range. This appears to be because the channel estimates obtained in the previous beam transmission cycles would not be reliable in the low SNR range; thus, the IMU sensor signals can compensate the degraded channel estimation.      
Note that, although both the proposed and  EKF methods perform the same measurement update step, the proposed method achieves a better performance owing to its more accurate prediction results in the prediction update step.  Note also that, although both our method and the method in \cite{MLAOA}  employ DNN for beam tracking, the proposed method performs better by leveraging the underlying domain knowledge in the measurement model.

\begin{figure} [t]  
	\centering   
	\subfigure[]{
    \includegraphics[width=95mm]{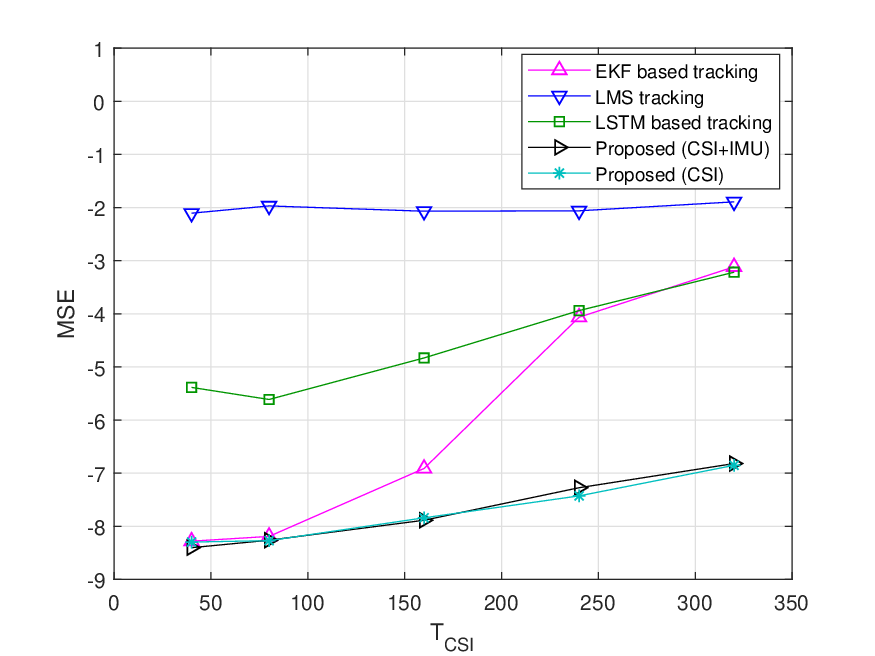}
   }
      \subfigure[]{        
        \includegraphics[width=95mm]{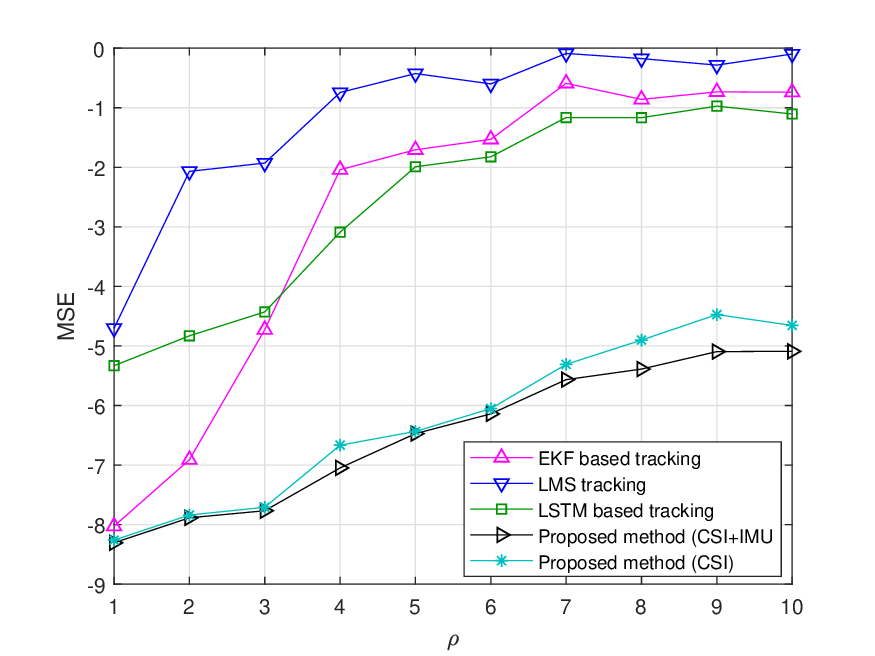}     
    }
\caption{Normalized MSE versus (a) $T_{CSI}$ and (b) $a_{avg}$. }   
\label{fig:mse_CSI}
\end{figure}

Fig.~\ref{fig:mse_SNR} shows the normalized MSE as a function of SNR for several beam tracking methods. The proposed method achieves a significant performance gain over the existing methods for all the cases considered. The performance gain of the proposed method also increases with $a_{avg}$. The proposed method can track rapidly varying channels better by using the DNN and IMU sensor measurements.

Fig.~\ref{fig:mse_CSI} illustrates the variation in the MSE performance with the beam transmission period $T_{CSI}$ and the angular velocity $a_{avg}$ when the SNR is set to $9$ dB. Fig.~\ref{fig:mse_CSI} (a) shows the plot of MSE versus $T_{CSI}$ when $a_{avg}$ is fixed to $0.2\pi$. As $T_{CSI}$ increases, the sounding beams are transmitted less frequently, and the beam tracking method experiences larger channel variations. With a small $T_{CSI}$, the performance of the EKF method is comparable to that of the proposed method. However, the performance of the EKF method severely deteriorates with $T_{CSI}$, and consequently, the performance gap between these two methods increases rapidly. Fig.~\ref{fig:mse_CSI} (b) shows the plot of MSE versus $a_{avg}$ when $T_{CSI}$ is fixed to 160 slots. The performance of the beam tracking algorithms degrades as the channel changes more dynamically owing to the fast motion of the UE.  As $a_{avg}$ increases, the EKF method does not perform well because the linear channel model used in the EKF method does not sufficiently capture the behavior of time-varying channels. In contrast, the proposed method successfully models the complex channel behavior for reliable beam tracking.

\section{Conclusions}\label{sec:conclusion}
In this paper, we proposed a deep learning-based beam tracking method for mmWave communication. The proposed beam tracking method was designed to track fast-varying AoD and AoA due the motion of the UE device. We employed the LSTM to model the channel variation and predict the future distribution of the channel state based on the sequence of the previous channel estimates and IMU sensor measurements.  Our method is based on a sequential Bayesian estimation framework, in which the prediction model yields the prior distribution of the channel in the prediction update step, and the predicted distribution is used to update the channel estimate in the measurement update step. Thus, our method is a hybrid approach in that we used both an LSTM-based channel model and an analytical measurement model for beam tracking. Our simulation results  showed that the proposed method achieved a significant performance gain over the EKF baseline and outperformed the existing beam tracking methods, especially in high-mobility scenarios.

\section{Appendix} 
 \subsection{Derivation of Jacobian Matrix} \label{appen:jaco}
The Jacobian matrix $\mathbf{O}_{t}$ is expressed as 
\begin{align*}
    \mathbf{O}_{t} = \left[\frac{\partial q(\mathbf{\gamma}_{t})}{\partial \theta^{(b)}_{1,t}},  \frac{\partial q(\mathbf{\gamma}_{t})}{\partial \theta^{(m)}_{1,t}},\cdots ,
 \frac{\partial q(\mathbf{\gamma}_{t})}{\partial \theta^{(b)}_{L,t}}, \frac{\partial q(\mathbf{\gamma}_{t})}{\partial \theta^{(m)}_{L,t}} \right],
\end{align*}
whose elements in the $(M(i-1)+j)$th row are given by
\begin{align*}
%\frac{\partial q^{(M(i-1)+j)}}{\partial \alpha_{l,t}} &= \frac{1}{n_t n_r} \frac{1-e^{\left(k_b N_b (\theta_{l,t}^{(b)}-\nu^{(b)}_{t,i})\right)}}{1-e^{\left(k_b (\theta_{l,t}^{(b)}-\nu^{(b)}_{t,i})\right)}}\frac{1-e^{\left(k_m N_m (\theta_{l,t}^{(m)}-\nu^{(m)}_{t,j})\right)}}{1-e^{\left(k_m (\theta_{l,t}^{(m)}-\nu^{(m)}_{t,j})\right)}}
\frac{\partial q^{(M(i-1)+j)}}{\partial \theta^{(b)}_{l,t}} &= \frac{\alpha_{l,t}}{n_t n_r} 
\frac{k_b e^{\left(k_b (\theta_{l,t}^{(b)}-\nu^{(b)}_{t,i})\right)}
- k_b N_b e^{\left(k_b N_b (\theta_{l,t}^{(b)}-\nu^{(b)}_{t,i})\right)}
- k_b (N_b-1) e^{\left(k_b (N_b+1) (\theta_{l,t}^{(b)}-\nu^{(b)}_{t,i})\right)}}
{\Big(1-e^{\left(k_b (\theta_{l,t}^{(b)}-\nu^{(b)}_{t,i})\right)}\Big)^2}
\\
& \times \frac{1-e^{\left(k_m N_m (\theta_{l,t}^{(m)}-\nu^{(m)}_{t,j})\right)}}
{1-e^{\left(k_m (\theta_{l,t}^{(m)}-\nu^{(m)}_{t,j})\right)}}
\\
\frac{\partial q^{(M(i-1)+j)}}{\partial \theta^{(m)}_{l,t}} &= \frac{\alpha_{l,t}}{n_t n_r} 
\frac{1-e^{\left(k_b N_b (\theta_{l,t}^{(b)}-\nu^{(b)}_{t,i})\right)}}
{1-e^{\left(k_b (\theta_{l,t}^{(b)}-\nu^{(b)}_{t,i})\right)}} \\ &\times
\frac{k_m e^{\left(k_m (\theta_{l,t}^{(m)}-\nu^{(m)}_{t,j})\right)}
- k_m N_m e^{\left(k_m N_m (\theta_{l,t}^{(m)}-\nu^{(m)}_{t,j})\right)}
- k_m (N_m-1) e^{\left(k_m (N_m+1) (\theta_{l,t}^{(m)}-\nu^{(m)}_{t,j})\right)}}
{\Big(1-e^{\left(k_m (\theta_{l,t}^{(m)}-\nu^{(m)}_{t,j})\right)}\Big)^2} ,  
\end{align*}    
where $ k_b = -\frac{j2\pi d_b}{\lambda}$ and $k_m = -\frac{j2\pi d_m}{\lambda}$.  
% The Jacobian matrix $\mathbf{O}_t$ is $MN \times 2L$ matrix, 

\bibliographystyle{IEEEbib}
\bibliography{ref.bib}
%\bibliography{IEEEabrv,CS_refs}
% that's all folks
\end{document}